\documentclass[12pt]{article}

\usepackage{amsmath,amssymb}
\usepackage{bm}
\usepackage{graphicx}
\usepackage{booktabs}
\usepackage{dcolumn}
\usepackage{microtype}
\usepackage[a4paper,margin=1in]{geometry}
\usepackage[hidelinks]{hyperref}
\usepackage{tikz}
\usetikzlibrary{arrows.meta,positioning}
\begin{document}

\title{Field-Free Transverse Aharonov--Bohm Phase Gate for an Orbital $l$-Qubit}

\author{Ju Gao\thanks{Corresponding author: jugao@illinois.edu} and Fang Shen\\[1mm]
\normalsize Department of Electrical and Computer Engineering,\\
\normalsize University of Illinois Urbana--Champaign, Urbana, Illinois 61801, USA}

\date{}

\maketitle
\begin{abstract}
The Aharonov--Bohm (AB) effect is usually read out through phase differences associated with spatially distinct electron paths. We show that confined orbital modes provide a same-path alternative: a core-confined magnetic flux writes opposite propagation phases on the co-propagating modes $|\pm l\rangle$ of a straight annular electron guide while the transported electron-wave support remains field free. In a spin-resolved Dirac treatment, the phase is carried by the overlap of the field-free vector potential $A_\phi$ with the mode's azimuthal conserved-current texture. The spin-dependent radial-gradient current becomes a boundary term that cancels when the complete finite-wall evanescent tail is retained, leaving the spin-independent orbital phase $\Delta\phi_{ln}\propto l\Phi L_{\rm int}\langle\rho^{-2}\rangle_{ln}/v_z$. The matched $|\pm l\rangle$ modes therefore realize a same-path $R_z(2\delta_l)$ gate, with differential internal-mode readout and common-mode phase rejection. For $a=75\,\mathrm{nm}$, $R=95\,\mathrm{nm}$, $L_{\rm int}=1\,\mathrm{mm}$, $E_z=10\,\mathrm{meV}$, and $|l|=10$, the gate angle is $2.315\,\mathrm{rad/G}$ and $R_z(\pi)$ occurs at $1.357\,\mathrm{G}$. Finite-barrier, mode-spacing, disorder-mismatch, and readout-visibility checks quantify the main implementation constraints. More broadly, the result connects a mode-resolved AB energy shift to a measurable propagation operation and shows how the spatially distributed conserved current of a Dirac wave can become an operational quantum-control resource.
\end{abstract}

\section{Introduction}

Orbital angular momentum provides a natural internal degree of freedom for coherent electron-wave control. Electron-vortex modes with phase winding $e^{il\phi}$ can be generated and sorted experimentally~\cite{UchidaTonomura2010,VerbeeckEtAl2010,McMorranEtAl2011,GrilloEtAl2017}, making opposite-winding modes $|\pm l\rangle$ a practical two-level basis. A useful orbital phase operation should act coherently on this internal mode pair while minimizing sensitivity to path imbalance. The magnetic Aharonov--Bohm (AB) effect supplies a field-free phase resource~\cite{AharonovBohm1959,Chambers1960,Tonomura1986,OlariuPopescu1985,PeshkinTonomura1989}, but conventional AB gate geometries encode the phase through spatial circulation or separated paths around the flux~\cite{PachosVedral2003,YuVoskoboynikov2008}. Here we ask whether a confined AB angular response can instead be converted into a same-path phase operation for a straight traveling orbital mode.

The key physical ingredient is the azimuthal current texture of an orbital eigenmode. Free-space AB Bessel modes are known to carry a flux-dependent azimuthal Schr\"odinger probability current and kinetic orbital angular momentum~\cite{BliokhEtAl2012}. Our earlier work derived the corresponding $l$-resolved AB coupling energy for stationary confined Dirac modes~\cite{GaoShen2026}. That result established a mode-resolved stationary coupling energy but did not address its accumulation in a traveling eigenmode or its use as an internal phase operation. The present work takes that next step: finite-wall confinement supplies a discrete normalizable transverse mode whose AB angular-energy shift accumulates reproducibly during longitudinal propagation, and the odd-in-$l$ response becomes a relative phase between matched $|+l\rangle$ and $|-l\rangle$ components. The same physics is thereby converted from a stationary energy response into a directly readable orbital-mode operation.

For the straight annular guide considered below, the electron-wave support excludes the flux-bearing core. The centroid propagates along $z$, while the field-free vector potential $A_\phi$ overlaps the mode's intrinsic azimuthal Dirac current. We refer to the resulting mode-resolved propagation channel as the transverse Aharonov--Bohm (TAB) phase because the current--potential projection is azimuthal while the centroid transport is longitudinal. A spin-resolved calculation shows that the radial-gradient current becomes a boundary term that cancels only after the complete finite-wall evanescent tail is retained; the surviving orbital term is spin independent to leading nonrelativistic order. The same result follows from the gauge-covariant angular Hamiltonian. Encoding $|\pm l\rangle$ as a qubit then gives an $R_z$ gate, an internal-mode analyzer provides differential readout with common-mode phase rejection, and angular-momentum selection suppresses direct opposite-winding conversion by low-order angular disorder at high $|l|$. The construction does not add a new interaction to the Dirac theory; rather, it retains the spatially distributed electron wave and its conserved current long enough for that local structure to appear as an experimentally addressable phase operation.

In the familiar closed-path form, the AB phase is
\begin{equation}
\Delta\phi_{\rm AB}
=
-\frac{e}{\hbar}
\oint_C \mathbf A\cdot d\mathbf r
=
-2\pi\frac{\Phi}{\Phi_0},
\quad
\Phi_0=\frac{h}{e}.
\label{eq:ABphase}
\end{equation}
A propagating confined mode $e^{ikz}$ carries a spatially extended Dirac conserved-current texture: its centroid propagates along $z$, whereas the field-free vector potential and the mode's azimuthal current are transverse to the propagation direction. To first order in the confined flux, the Dirac minimal-coupling perturbation $\hat H_A=ec\,\boldsymbol{\alpha}\cdot\mathbf A$ gives $\Delta E^{(1)}=-\int_{\mathcal V_e}\mathbf j\cdot\mathbf A\,dV$. Propagation for a time $T$ therefore accumulates
\begin{equation}
\begin{aligned}
\Delta\phi_{\rm TAB}
&=\frac{1}{\hbar}\int dt\int_{\mathcal V_e}
j_\phi A_\phi\,dV,\\
\mathbf A\cdot d\mathbf r_c
&=A_z\,dz=0,
\quad j_\phi A_\phi\not\equiv0,
\end{aligned}
\label{eq:TABdefinition}
\end{equation}
where $\mathcal V_e$ denotes the electron-wave support within the interaction section. Equation~\eqref{eq:TABdefinition} is a current representation of the same gauge-covariant AB response derived below from the angular Hamiltonian; the observable is the complete relative phase between matched $|\pm l\rangle$ modes, not the pointwise factor $j_\phi A_\phi$.

Under a single-valued, time-independent gauge transformation on the electron-support region, $\mathbf A\rightarrow\mathbf A+\nabla\chi$, and suppressing the common time integral for the stationary transverse mode,
\[
\int_{\mathcal V_e}\mathbf j\cdot\nabla\chi\,dV
=
\oint_{\partial\mathcal V_e}\chi\,\mathbf j\cdot d\mathbf S
-
\int_{\mathcal V_e}\chi\,\nabla\cdot\mathbf j\,dV .
\]
The bulk term vanishes by $\nabla\cdot\mathbf j=0$. The radial surface contribution vanishes because the stationary annular mode carries no normal radial current and its evanescent tail decays at infinity. The longitudinal end-face contribution is identical for the matched $|\pm l\rangle$ modes, which share the same $j_z$, and therefore cancels from their internal-interference phase. The relative phase is thus gauge invariant.

\begin{figure}[t]
\centering
\includegraphics[width=0.45\textwidth]{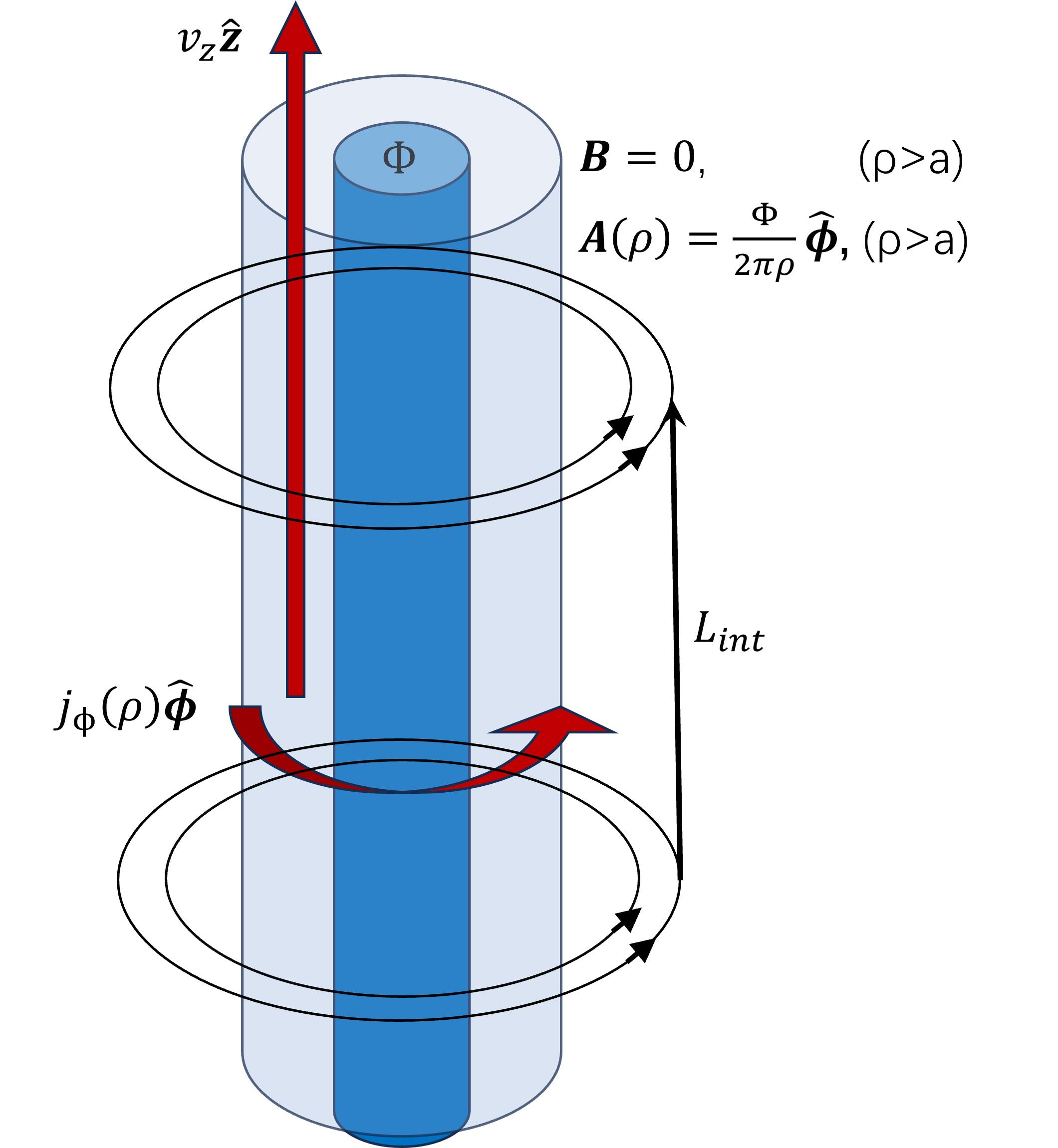}
\caption{Core-excluded same-path AB section. The flux-bearing core lies outside the support of the traveling electron wave. The centroid propagates along $z$ with no path projection onto $\mathbf A$, while the mode's azimuthal current $j_\phi$ projects onto $A_\phi$ over the interaction length $L_{\rm int}$.}
\label{fig:annulus}
\end{figure}

Figure~\ref{fig:annulus} shows the device geometry. A coaxial inner wall excludes the flux-bearing core from the electron-wave support, while cylindrical confinement supplies eigenmodes with definite radial structure and orbital quantum number $l$. Propagation through an interaction length $L_{\rm int}$ accumulates the phase over $T=L_{\rm int}/v_z$. A gated or electrostatic annulus guides the eigenmode, a shielded core supplies flux, and short mode converters prepare and analyze the $\pm l$ components. The core may extend through the assembly, so $L_{\rm int}$ is the calibrated separation between the preparation and analyzer sections rather than a spatial turn-on of the flux. The two logical components therefore remain in the same straight guide, eliminating arm imbalance while allowing shared longitudinal phase to be rejected by internal-mode differential readout.

\section{Confined Dirac Mode and Same-Path AB Phase}

We now derive the phase written by the field-free $A_\phi$ onto a matched annular Dirac eigenmode. The annular confinement is
\begin{equation}
U(\rho)=
\begin{cases}
\infty, & \rho<a,\\
0, & a<\rho<R,\\
U_0, & \rho>R,
\end{cases}
\label{eq:potential}
\end{equation}
where the finite outer wall admits an evanescent tail. Let the leading upper-component radial envelope be
\begin{equation}
u_{ln}(\rho)=
\begin{cases}
F_{ln}(\rho), & a<\rho<R,\\
\kappa_{ln}K_l(\xi_{ln}\rho), & \rho>R,
\end{cases}
\label{eq:radial}
\end{equation}
with
\begin{equation}
F_{ln}(\rho)
=
J_l(\zeta_{ln}\rho)Y_l(\zeta_{ln}a)
-
Y_l(\zeta_{ln}\rho)J_l(\zeta_{ln}a).
\label{eq:Fl}
\end{equation}
Thus $u_{ln}(a)=0$, and the electron wave has no support in the
flux-bearing core. Here $\zeta_{ln}$ and $\xi_{ln}$ are the oscillatory and evanescent radial wave numbers, respectively, and $\kappa_{ln}$ is the exterior matching amplitude. For the finite outer barrier, the modified Bessel function $K_l(\xi_{ln}\rho)$ describes the normalizable evanescent tail for $\rho>R$. Continuity of the radial envelope and its derivative gives
\begin{equation}
\begin{aligned}
F_{ln}(R)
&=
\kappa_{ln}K_l(\xi_{ln}R),\\
F_{ln}'(R)
&=
\left.
\partial_\rho\!\left[
\kappa_{ln}K_l(\xi_{ln}\rho)
\right]\right|_{\rho=R}.
\end{aligned}
\label{eq:matching}
\end{equation}
These conditions determine the finite-wall eigenmode and extend the
radial density continuously into the decaying exterior sector, consistent
with the finite-wall Dirac-current continuation studied
previously~\cite{GaoShen2024Evanescent}. Retaining this evanescent tail is
essential below, where it completes the boundary cancellation of the
radial-gradient contribution. 

To leading nonrelativistic order, the positive-energy Dirac spin-up traveling state is
\begin{equation}
\psi^{\uparrow}_{lnk}
=N_{ln}e^{il\phi}e^{ikz}
\begin{pmatrix}
u_{ln}\\
0\\
\eta k u_{ln}\\
-i\eta e^{i\phi}D_lu_{ln}
\end{pmatrix}.
\label{eq:spinor}
\end{equation}
The factor $e^{ikz}$ makes the transverse eigenmode a traveling wave and supplies $T=L_{\rm int}/v_z$. Here
\[
\begin{gathered}
D_l=\partial_\rho-l/\rho,\quad \eta=\hbar/(2mc).
\end{gathered}
\]

Define the normalized radial density
\[ n_{ln}(\rho)\equiv N_{ln}^{2}u_{ln}^{2}(\rho),
\quad
2\pi\int_a^\infty \rho\,n_{ln}(\rho)\,d\rho=1+O(v^2/c^2).
\]
Using $\alpha_\phi=-\alpha_x\sin\phi+\alpha_y\cos\phi$, direct contraction of Eq.~\eqref{eq:spinor} and its spin-down partner with $\mathbf j=-ec\,\psi^\dagger\boldsymbol{\alpha}\psi$ gives the longitudinal and azimuthal current components
\begin{equation}
\begin{aligned}
j_z
&=-\frac{e\hbar k}{m}n_{ln},\\
j_\phi^{(s)}
&=s\,\frac{e\hbar}{2m}\frac{dn_{ln}}{d\rho}
-\frac{e\hbar l}{m\rho}n_{ln},
\quad s=\pm1.
\end{aligned}
\label{eq:current}
\end{equation}
Here $s=+1$ ($-1$) denotes spin up (down). The radial-gradient contribution reverses under spin reversal, whereas the orbital contribution is unchanged, proportional to $l$, and changes sign under $l\rightarrow-l$.

For a flux confined to the inaccessible core,
\begin{equation}
\mathbf B=0, \quad
A_\phi(\rho)=\frac{\Phi}{2\pi\rho},
\quad \rho>a, 
\label{eq:Aphi}
\end{equation}
and a uniform straight section of duration $T=L_{\rm int}/v_z$, Eq.~\eqref{eq:TABdefinition} reduces to
\begin{equation}
\Delta\phi_{ln}
=\frac{\Phi T}{\hbar}\int_a^\infty j_\phi(\rho)\,d\rho.
\label{eq:phaseintegral}
\end{equation}
Substitution of Eq.~\eqref{eq:current} yields
\begin{equation}
\Delta\phi_{ln}^{(s)}
=s\,\frac{e\Phi T}{2m}
\bigl[n_{ln}(\infty)-n_{ln}(a)\bigr]
-\frac{e l\Phi T}{m}  \!\!
\int_a^\infty \!\! \frac{n_{ln}(\rho)}{\rho}\,d\rho.
\label{eq:phasefull}
\end{equation}

The radial-gradient current, which reverses under spin reversal, has become a complete boundary contribution. It vanishes because the normalizable finite-wall mode satisfies
$n_{ln}(a)=n_{ln}(\infty)=0$; retaining the evanescent sector is
essential to complete this cancellation. The orbital term is unchanged between the two spin partners and survives. The TAB
phase is therefore spin independent to this order,
\begin{equation}
\Delta\phi_{ln}
=
-\frac{\hbar}{m}\,
l\frac{\Phi}{\Phi_0}
\frac{L_{\rm int}}{v_z}
\left\langle\rho^{-2}\right\rangle_{ln},
\label{eq:TABphase}
\end{equation}
where the inverse-square radial expectation value of the $(l,n)$ mode is
\begin{equation}
\left\langle\rho^{-2}\right\rangle_{ln}
=
2\pi\int_a^\infty
\frac{n_{ln}(\rho)}{\rho}\,d\rho.
\label{eq:rho2}
\end{equation}

Equation~\eqref{eq:TABphase} gives the explicit mode-resolved TAB
phase: it is linear in $l$ and $\Phi$ to first order, accumulates over $L_{\rm int}/v_z$, and is nonzero for $l\neq0$ although the centroid path has no projection onto $\mathbf A$. A complementary
gauge-covariant formulation makes the same flux dependence explicit. To the same leading nonrelativistic order, throughout the electron-support region,
\begin{equation}
\hat H_\phi(\Phi)
=
\frac{\left(
-i\hbar\partial_\phi+e\rho A_\phi
\right)^2}{2m\rho^2}=
\frac{1}{2m\rho^2}
\left(
-i\hbar\partial_\phi+\hbar\frac{\Phi}{\Phi_0}
\right)^2.
\label{eq:covariantH}
\end{equation}
For the normalized zero-flux transverse mode $\varphi_{ln}$ with radial envelope $u_{ln}(\rho)$ and angular factor $e^{il\phi}$, the flux-induced angular-energy expectation-value shift is

\begin{equation}
\begin{aligned}
\Delta E_{ln}(\Phi)
&=
\left\langle\varphi_{ln}\left|
\hat H_\phi(\Phi)-\hat H_\phi(0)
\right|\varphi_{ln}\right\rangle
\\
&=
\frac{\hbar^2}{2m}
\left[
\left(l+\frac{\Phi}{\Phi_0}\right)^2-l^2
\right]
\left\langle\rho^{-2}\right\rangle_{ln},
\\
\Delta E_{ln}^{(1)}
&=
\frac{\hbar^2}{m}\,
l\frac{\Phi}{\Phi_0}
\left\langle\rho^{-2}\right\rangle_{ln}.
\end{aligned}
\label{eq:covariantenergy}
\end{equation}
The accumulated phase
$-\Delta E_{ln}^{(1)}L_{\rm int}/(\hbar v_z)$ reproduces
Eq.~\eqref{eq:TABphase}; equivalently, the first derivative of the gauge-covariant angular Hamiltonian at $\Phi=0$ gives the current-projection result through the Hellmann--Feynman relation. In contrast to the free-space AB Bessel modes of Ref.~\cite{BliokhEtAl2012}, the confined annular state has a discrete, normalizable transverse profile with a definite
$\langle\rho^{-2}\rangle_{ln}$; its flux-dependent angular energy therefore writes a reproducible phase during guided propagation over $L_{\rm int}/v_z$. The distributed-current projection and the gauge-covariant angular Hamiltonian are equivalent descriptions of the same first-order TAB operation. The term quadratic in $\Phi$ is even in $l$ and therefore common to a matched $|\pm l\rangle$ pair. Thus, to first order, the mode-resolved phase reverses under either $l\rightarrow-l$ or $\Phi\rightarrow-\Phi$, remains unchanged under spin reversal, and vanishes for $l=0$.

\section{Orbital-Qubit Phase Gate and Readout}

The odd-in-$l$ phase provides a direct logical operation on opposite-winding modes. This structure selects the opposite phase-winding modes as a two-level phase basis:
\begin{equation}
|0_l\rangle\equiv|+l,n,k\rangle,
\quad
|1_l\rangle\equiv|-l,n,k\rangle.
\label{eq:lqubit}
\end{equation}
These are opposite $e^{\pm il\phi}$ phase-winding and current configurations of the same guided transverse mode family. Their coherent superposition remains in one spatial guide, so the logical phase is written between internal orbital components rather than between separated interferometer arms. The magnitude $|l|$ is a phase-response knob, while its sign reverses the phase without changing the matched radial or longitudinal environment. With
$\delta_l=|\Delta\phi_{+l,n}|$ and
$\Delta\phi_{\pm l,n}=\mp\delta_l$, the same-path AB section transforms an
arbitrary logical state according to
\begin{equation}
c_0|0_l\rangle+c_1|1_l\rangle
\longrightarrow
c_0e^{-i\delta_l}|0_l\rangle
+c_1e^{+i\delta_l}|1_l\rangle.
\label{eq:stateoperation}
\end{equation}
Therefore, the field-free current--potential coupling implements the unitary phase transformation
\begin{equation}
U_{\rm TAB}
=
\begin{pmatrix}
e^{-i\delta_l}&0\\
0&e^{+i\delta_l}
\end{pmatrix}
=
R_z(2\delta_l).
\label{eq:unitary}
\end{equation}
Thus the same-path AB section acts diagonally in the logical basis, preserving
the populations while writing the controllable relative phase
$2\delta_l$.

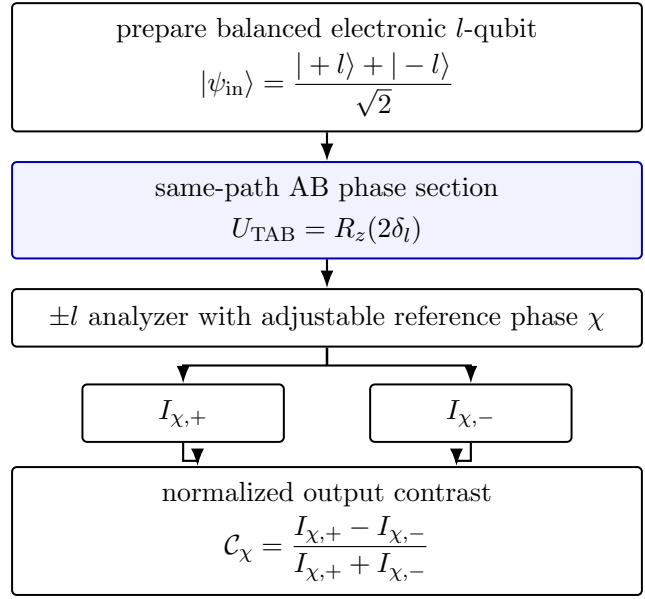
\begin{figure}[t]
\centering
\begin{tikzpicture}[
    scale=0.95,
    transform shape,
    >=Latex,
    node distance=4.0mm,
    box/.style={
        draw,
        rounded corners=1.8pt,
        align=center,
        minimum height=8.0mm,
        inner xsep=4.5mm,
        inner ysep=2.0mm,
        font=\small,
        line width=0.75pt
    },
    mainbox/.style={
        box,
        minimum width=0.55\textwidth
    },
    flow/.style={
        ->,
        line width=0.8pt
    }
]

\node[mainbox] (prep)
{prepare balanced electronic \(l\)-qubit\\[1mm]
\(\displaystyle
|\psi_{\rm in}\rangle
=
\frac{|+l\rangle+|-l\rangle}{\sqrt{2}}
\)};

\node[
    mainbox,
    below=of prep,
    draw=blue!70!black,
    fill=blue!5
] (tab)
{same-path AB phase section\\[1mm]
\(\displaystyle
U_{\rm TAB}
=
R_z(2\delta_l)
\)};

\node[mainbox, below=of tab] (ana)
{\(\pm l\) analyzer with adjustable reference phase \(\chi\)};

\node[
    box,
    below=5.0mm of ana,
    xshift=-20mm,
    minimum width=28mm,
    minimum height=8mm
] (ip)
{\(\displaystyle I_{\chi,+}\)};

\node[
    box,
    below=5.0mm of ana,
    xshift=20mm,
    minimum width=28mm,
    minimum height=8mm
] (im)
{\(\displaystyle I_{\chi,-}\)};

\node[
    mainbox,
    below=16.5mm of ana,
    minimum height=12mm
] (contrast)
{normalized output contrast\\[1mm]
\(\displaystyle
\mathcal C_\chi
=
\frac{I_{\chi,+}-I_{\chi,-}}
     {I_{\chi,+}+I_{\chi,-}}
\)};

\draw[flow] (prep) -- (tab);
\draw[flow] (tab) -- (ana);

\coordinate (split) at ([yshift=-2.5mm]ana.south);
\draw[line width=0.95pt] (ana.south) -- (split);
\draw[flow] (split) -| (ip.north);
\draw[flow] (split) -| (im.north);

\draw[flow]
(ip.south) -- ++(0,-2.5mm)
-| ([xshift=-18mm]contrast.north);

\draw[flow]
(im.south) -- ++(0,-2.5mm)
-| ([xshift=18mm]contrast.north);

\end{tikzpicture}

\caption{
Same-path preparation, phase transformation, and readout of the electronic
\(l\)-qubit. The field-free transverse Aharonov--Bohm section writes the
relative phase \(2\delta_l\) between the logical states
\(|+l\rangle\) and \(|-l\rangle\). An adjustable \(\pm l\) analyzer converts
this phase into the normalized output contrast \(\mathcal C_\chi\).
}
\label{fig:readout}
\end{figure}

To turn the orbital phase operation into an experimentally executable
measurement, we convert the written relative phase into a differential
readout through internal interference of the two logical components. As
shown in Fig.~\ref{fig:readout}, a balanced input
$(|+l\rangle+|-l\rangle)/\sqrt{2}$ traverses the common-path AB
section, acquires the opposite phases $\mp\delta_l$, and therefore
emerges with the relative phase $2\delta_l$. An analyzer with adjustable
reference phase $\chi$ then projects the two components onto
interfering output channels. Let $I_{\chi,\pm}$ denote the corresponding
detector currents or accumulated event rates, with
$I_0=I_{\chi,+}+I_{\chi,-}$:
\begin{equation}
I_{\chi,\pm}
=
\frac{I_0}{2}
\left[
1\pm\cos(2\delta_l-\chi)
\right].
\label{eq:readoutintensity}
\end{equation}
The phase is therefore read directly as the normalized differential
contrast
\begin{equation}
\mathcal C_\chi
\equiv
\frac{I_{\chi,+}-I_{\chi,-}}
{I_{\chi,+}+I_{\chi,-}}
=
\cos(2\delta_l-\chi).
\label{eq:contrast}
\end{equation}
Equations~\eqref{eq:readoutintensity} and \eqref{eq:contrast} describe ideal balanced preparation and mode projection. Preparation and analyzer imperfections can be represented phenomenologically by an interference visibility $\mathcal V$. For symmetric imperfections that reduce modal coherence or overlap without introducing a channel imbalance,
\begin{equation}
\mathcal C^{\rm obs}_\chi
=
\mathcal V\cos(2\delta_l-\chi),
\qquad 0\leq\mathcal V\leq1.
\label{eq:visibility}
\end{equation}
Imperfect mode overlap then reduces the contrast amplitude without shifting the extracted phase. Unequal channel efficiencies or asymmetric modal crosstalk can additionally generate an offset or phase bias and must be calibrated in an implementation. At quadrature, $\chi=\pi/2$,
\begin{equation}
\mathcal C^{\rm obs}_{\pi/2}
=
\mathcal V\sin(2\delta_l)
\simeq
2\mathcal V\delta_l.
\label{eq:quadrature}
\end{equation}
Thus the same-path interferometric readout converts the written AB phase into a directly measurable differential signal, with the opposite mode phases doubling the small-signal response. If both logical components acquire an additional common phase $\phi_{\rm c}$, the output state is multiplied only by the global factor $e^{i\phi_{\rm c}}$, which cancels from $\mathcal C_\chi$. The readout therefore rejects common-mode phase noise, including shared longitudinal dynamical phase and common path-length drift, while retaining the odd-in-$l$ orbital phase. Flux noise, $l$-odd perturbations, and other differential mode shifts remain.

\section{Device Scale and Mode-Mixing Robustness}

As a representative device-scale benchmark, we take
$a=75\,\mathrm{nm}$, $R=95\,\mathrm{nm}$, $|l|=10$, $n=1$,
$E_z=10\,\mathrm{meV}$, and $L_{\rm int}=1\,\mathrm{mm}$. The
$75\,\mathrm{nm}$ inner radius places the device at a
sub-$100\,\mathrm{nm}$ lateral scale while retaining a
$20\,\mathrm{nm}$ annular width. Equation~\eqref{eq:TABphase} shows that
the phase can be enhanced by increasing $|l|$ or the enclosed flux,
extending the interaction length, reducing the longitudinal velocity, or
tightening the radial confinement to increase
$\langle\rho^{-2}\rangle_{ln}$, subject to coherent mode transport and
core-field exclusion. For numerical transparency, we evaluate the radial
mode in the hard-wall limit $U_0\rightarrow\infty$, for which the roots
satisfy
\begin{equation}
J_l\!\left(\frac{R}{a}x\right)Y_l(x)
-Y_l\!\left(\frac{R}{a}x\right)J_l(x)=0,
\qquad x=\zeta_{ln}a.
\label{eq:hardwallroot}
\end{equation}
For $R/a=95/75$, the lowest $|l|=10$ root is
$x_{10,1}=14.7231081$. This hard-wall approximation is used only to set
the benchmark values; the finite-wall derivation above establishes that
the boundary closure underlying the TAB phase remains intact when the
evanescent tail is retained. As a finite-barrier check, solving the
matching condition in Eq.~\eqref{eq:matching} for $U_0=1$ and
$5\,\mathrm{eV}$ gives
$\langle\rho^{-2}\rangle_{10,1}=1.38469\times10^{14}$ and
$1.38645\times10^{14}\,\mathrm{m}^{-2}$, respectively, compared with
$1.38788\times10^{14}\,\mathrm{m}^{-2}$ in the hard-wall limit. Thus the
hard-wall radial weight differs by only $0.23\%$ at $1\,\mathrm{eV}$ and
$0.10\%$ at $5\,\mathrm{eV}$; the corresponding phase sensitivity is
$2.3098$ and $2.3127\,\mathrm{rad/G}$, compared with
$2.3151\,\mathrm{rad/G}$ in the hard-wall benchmark.

\begin{table}[t]
\caption{
Hard-wall annular benchmark for the orbital $l$-qubit phase gate.
Here $x_{ln}=\zeta_{ln}a$, $B_G=B/(1\,\mathrm{G})$, and
$\theta_z=2\delta_l$.
}
\label{tab:benchmark}

\footnotesize
\setlength{\tabcolsep}{1.0pt}
\renewcommand{\arraystretch}{1.02}

\resizebox{0.96\columnwidth}{!}{%
\begin{tabular}{@{}l l@{}}
\toprule
Quantity & Value \\
\midrule

Geometry &
$a=75\,\mathrm{nm},\ R=95\,\mathrm{nm}$ \\

Interaction &
$L_{\rm int}=1\,\mathrm{mm},\ T=16.861\,\mathrm{ns}$ \\

Mode &
$l=\pm10,\ n=1,\ E_z=10\,\mathrm{meV}$ \\

Radial solution &
$x_{10,1}=14.7231081,\quad \epsilon_\perp=1.46825\,\mathrm{meV}$ \\

Radial weight &
$\langle\rho^{-2}\rangle_{10,1}
=1.38788\times10^{14}\,\mathrm{m}^{-2}$ \\

Radial gap &
$E_{10,2}-E_{10,1}=2.82451\,\mathrm{meV}$ \\

$l=9,11$ gaps &
$E_{10,1}-E_{9,1}=100.49\,\mu\mathrm{eV}$ \\
& $E_{11,1}-E_{10,1}=111.01\,\mu\mathrm{eV}$ \\

Longitudinal velocity &
$v_z=5.93097\times10^4\,\mathrm{m\,s^{-1}}$ \\

Flux ratio &
$\Phi/\Phi_0=4.27294\times10^{-4}B_G$ \\

Mode phase &
$\Delta\phi_{\pm10}
=\mp1.15755B_G\,\mathrm{rad}$ \\

Gate angle &
$\theta_z=2.31510B_G\,\mathrm{rad}$ \\

Phase sensitivity &
$\partial\theta_z/\partial B
=2.31510\,\mathrm{rad/G}$ \\

Ideal quadrature contrast ($\mathcal V=1$, $1\,\mathrm{mG}$) &
$\mathcal C_{\pi/2}\simeq2.315\times10^{-3}$ \\

$R_z(\pi)$ operation &
$B_\pi=1.35700\,\mathrm{G}$ \\

Flux at $\pi$ gate &
$\Phi_\pi/\Phi_0=5.79839\times10^{-4}$ \\
\bottomrule
\end{tabular}
}
\end{table}

Table~\ref{tab:benchmark} shows that a $1\,\mathrm{mm}$ same-path AB
section produces a gate angle of $2.315\,\mathrm{rad/G}$ and reaches
$R_z(\pi)$ at $1.357\,\mathrm{G}$. At the quadrature operating point
$\chi=\pi/2$, a $1\,\mathrm{mG}$ field change produces an ideal
($\mathcal V=1$) normalized differential contrast of
$2.315\times10^{-3}$. The same calculation gives a radial excitation gap
of $2.825\,\mathrm{meV}$, while the nearest angular modes lie
$100.5\,\mu\mathrm{eV}$ below ($l=9$) and $111.0\,\mu\mathrm{eV}$ above
($l=11$) the $l=10$ mode. Thus radial leakage is substantially more
detuned than neighboring angular-mode leakage in this geometry.

At $E_z=10\,\mathrm{meV}$, the $1\,\mathrm{mm}$ interaction section
corresponds to a transit time $T=16.861\,\mathrm{ns}$. The benchmark
therefore requires preservation of the relative $|\pm10\rangle$ phase
over at least this propagation length and time. The achievable orbital
coherence length and time are implementation dependent and constitute
experimental requirements rather than assumptions of the phase
calculation. Candidate embodiments include vacuum or electrostatically
guided electron modes and solid-state annular channels; the excluded
flux core could be supplied by a shielded magnetic nanowire or a
superconducting flux-line structure, provided the transported electron
wave remains in a region of negligible magnetic field. Experimental
realization will additionally require high-purity $|\pm10\rangle$
preparation, mode-preserving transport, control of residual flux leakage
and differential disorder, and a calibrated $\pm l$ analyzer.

The high-$|l|$ encoding specifically suppresses direct opposite-winding
logical conversion, rather than all mode leakage. For an angular
perturbation
\begin{equation}
\begin{aligned}
V(\rho,\phi)
&=
\sum_m V_m(\rho)e^{im\phi},\\
\langle +l|V|-l\rangle
&\propto V_{2l},
\quad
\langle -l|V|+l\rangle
\propto V_{-2l},
\end{aligned}
\label{eq:mixingselection}
\end{equation}
Equation~\eqref{eq:mixingselection} shows that direct first-order conversion between the two logical states
requires the angular harmonic $m=\pm2l$. The $|\pm10\rangle$ pair therefore
requires $m=\pm20$ for a direct logical flip. Low-order disorder can still
couple $l=10$ to nearby orbital modes: $m=\pm1$ connects to $l=9,11$.
For the benchmark gaps in Table~\ref{tab:benchmark}, static elastic
scattering at $E_z=10\,\mathrm{meV}$ corresponds, to first order, to
longitudinal wave-number mismatches
$|\Delta k_z|\simeq2.57\times10^6\,\mathrm{m}^{-1}$ and
$2.84\times10^6\,\mathrm{m}^{-1}$, or mismatch lengths
$|\Delta k_z|^{-1}\simeq0.39\,\mu\mathrm{m}$ and
$0.35\,\mu\mathrm{m}$. Longitudinally smooth disorder is therefore
phase mismatched over the $1\,\mathrm{mm}$ interaction section, whereas
short-range roughness and other differential mode shifts remain relevant.
Increasing $|l|$ retains the useful direct-conversion selection rule while
also increasing the AB phase response linearly.

\section{Conclusion}

We have converted the Aharonov--Bohm angular response of a confined electron mode into a same-path orbital-qubit phase operation. In the spin-resolved Dirac-current description, the finite-wall evanescent tail closes the spin-dependent radial-gradient contribution as a boundary term, while the orbital current produces the gauge-invariant phase $\Delta\phi_{ln}\propto l\Phi L_{\rm int}\langle\rho^{-2}\rangle_{ln}/v_z$. The equivalent gauge-covariant angular Hamiltonian gives the same result. A matched $|\pm l\rangle$ pair therefore realizes $R_z(2\delta_l)$ without spatially separated logical paths, and an internal-mode analyzer converts the written phase into differential contrast with common-mode phase rejection. Increasing $|l|$ simultaneously increases the phase response and pushes direct first-order opposite-winding conversion to the angular harmonic $m=\pm2l$; neighboring-mode leakage remains governed by lower-order disorder and its longitudinal phase mismatch. The $l$- and $\Phi$-reversal signatures, spin independence to this order, and $l=0$ null provide direct checks of the operation. The result is the operational continuation of the previously derived mode-resolved AB coupling energy: a stationary current--potential response becomes a propagation phase with a defined input, control parameter, and readout. In this sense, the spatial structure of the Dirac wave is not only part of the state description; through its conserved-current texture it sets a measurable device-level operation. The annular geometry thus supplies a concrete bridge from confined Aharonov--Bohm mode physics to coherent orbital-electron control.

\section*{AUTHOR DECLARATIONS}

\subsection*{Conflict of Interest}
The authors have no conflicts to disclose.

\subsection*{Author Contributions}
Ju Gao: Conceptualization (equal); Investigation (equal); Visualization (equal); Writing -- original draft (lead). Fang Shen: Conceptualization (equal); Investigation (equal); Visualization (equal); Writing -- review \& editing (lead).

\section*{Data availability statement}
The data that support the findings of this study are available within the article.

\bibliographystyle{iopart-num}
\bibliography{TAB}

\end{document}